# Cooperative Spectrum Sensing over Generalized Fading Channels Based on Energy Detection

*He Huang, et.al.*

(Beijing University of Posts and Telecommunications, Beijing 100876, China)

**Abstract:** This paper analyzes the unified performance of energy detection (ED) of spectrum sensing (SS) over generalized fading channels in cognitive radio (CR) networks. The detective performance of SS schemes will be obviously affected by fading channel among communication nodes, and ED has the advantages of fast implementation, no requirement of priori received information and low complexity, so it is meaningful to investigate ED that is performed over fading channels such as Nakagami-$m$ channel and Rice channel, or generalized fading channels such as $\kappa$-$\mu$ fading distribution and $\eta$-$\mu$ fading distribution. The $\alpha$-$\kappa$-$\mu$ fading distribution is a generalized fading model that represents the nonlinear and small-scale variation of fading channels. The probability density function (*p.d.f.*) of instantaneous signal-to-ratio (SNR) of $\alpha$-$\kappa$-$\mu$ distribution is derived from the *p.d.f.* of envelope to evaluate energy efficiency for sensing systems. Next, the detection probability of detective model with Marcum-$Q$ function has been derived and the close-form expressions with moment generating function (*MGF*) method are deduced to achieve SS. Furthermore, novel and exact closed-form analytic expressions for average area under the receiver operating characteristics curve ($\overline{AUC}$) also have been deduced to analyze the performance characteristics of ED over $\alpha$-$\kappa$-$\mu$ fading channels. Besides, cooperative spectrum sensing (CSS) with diversity reception has been applied to improve the detection accuracy and mitigate the shadowed fading features with OR-rule. At last, the results show that the detection capacity can be evidently affected by $\alpha$-$\kappa$-$\mu$ fading conditions, but appropriate channel parameters will improve sensing performance. On the other hand, the established ED-fading pattern is approved by simulations under diversity combination with CSS at the receiving end and it can significantly enhance the detection performance of proposed algorithms.

## I. INTRODUCTION

Nowadays, as wireless services rapidly develop and spectrum wideband has been applied extensively, spectrum scarcity has been extremely serious because of sustained growth of data rate applications in wireless communications, based on this, cognitive radio (CR) networks have been proposed to effectively manage spectrum resources and efficiently make full use of limited spectrum bands [1-5]. Spectrum sensing (SS) is the most important part of CR technologies which enables secondary users (SUs) better access the allocated wideband of primary users (PUs) [6-8]. There are mainly three categories spectrum sensing techniques, which include matched filter detection, cyclostationary feature detection and energy detection (ED). The matched filter detection technique that is normally implemented in the digital domain needs exact bandwidth and modulation type transmission information, but it requires the minimum possible number of samples since match filter uses the optimal processing [9,

10]. Similarly, statistical characteristics of transmitted signals are applied to cyclostationary feature detection to improve the detection probability of detective model [11]. Moreover, ED is the most popular non-coherent signal detection algorithm and it has the advantages in low-complexity, implementation simplicity and detection without a priori knowledge [12], it employs ED radiometer at the receiver to compare the received energy value with fixed threshold and determine the state of PU is absent or present instantaneously. Therefore, considering the detection performance of ED algorithm at the real time is a highly significant research work [12-15].

Urkowitz firstly adopted binary hypothesis-testing signal detection over a flat band-limited Gaussian noise channel by deriving the probability of detection $P_d$ and false alarm $P_f$ which follow the central chi-square and non-central chi-square distribution respectively [16]. Next, Kostylev and Alouini et al. considered ED model over conventional fading channels and even multipath fading conditions such as Rayleigh, Nakagami$q$ and Nakagami$m$ [17-19]. Since then, ED algorithm has been widely used in corresponding fading scenarios with diversity combining, for example, maximal ratio combining (MRC), selection combining (SC) and equal gain combining (EGC) etc [20-23]. Numerous studies on the basis of ED are investigated to achieve SS over different communication circumstances, in Ref. [24] unified close-form expressions of average detection probability of ED with cooperative spectrum sensing (CSS) are derived over generalized multipath fading channels to improve the efficiency and usefulness of SS. In Ref. [25] novel expressions have been derived over extended generalized $K$ composite fading channels to provide an unified model for wireless communication channel statistics. Likewise, exact closed-form expressions of detection pattern over $N$*Rayleigh channels are derived with Meijer G-function and Marcum-$Q$ function, then it is extended to the case of square-law selection (SLS) to achieve better performance than conventional detection over Rayleigh fading conditions.

On the other hand, as wireless radio propagation is affected by multipath and shadowing effects simultaneously, the adequate fading expressions are needed to represent the fundamental characteristics for composite statistical patterns [26]. Hence, the generalized fading channels $\kappa$-$\mu$ and $\eta$-$\mu$ which describe the line-of-sight (LOS) and non-line-of-sight (NLOS) communication conditions are proposed to provide accurate representation of radio propagation [27]. Next, the fading distributions $\alpha$-$\kappa$-$\mu$ and $\alpha$-$\eta$-$\mu$, are derived to represent the non-linear LOS and NLOS small-scale variation of the fading signals, besides the performance evaluation of digital communication systems and the capacity analysis have been considered over $\alpha$-$\eta$-$\mu$ fading channels [28, 29]. Then a general and comprehensive complex fading model $\alpha$-$\eta$-$\kappa$-$\mu$ is put forward to account for short-term propagation phenomena by employing joint phase-envelop method to represent the device-to-device communications, vehicle-to-vehicle communications and indoor-to-outdoor propagation in 5G [28]. Moreover, severe fading channels and composite fading/shadowed models such as $\kappa$-$\mu$ extreme distribution, $\kappa$-$\mu$/Gamma, $\eta$-$\mu$/Gamma and $K$-distribution have been developed to measure the practical communication channels in the complex conditions [31-36].

The $\alpha$-$\kappa$-$\mu$ distribution is a very general and flexible fading model which contains more fading parameters to describe the nonlinear LOS small-scale transmission scenario.

Specially, it contains $\alpha$-$\mu$ when $\kappa$ is approached to 0 and $\kappa$-$\mu$ distribution when α is 2. Furthermore, the special classical distributions such as Rayleigh ($\alpha=2$, $\kappa=0$, $\mu=1$), Rice ($\alpha=2$, $\kappa=k$, $\mu=1$), Nakagami-m ($\alpha=2$, $\kappa=0$, $\mu=m$) and One-sided Gaussian distribution ($\alpha=2$, $\kappa=0$, $\mu=0.5$) can be obtained by estimating the fading parameters $\alpha$, $\kappa$ and $\mu$ [37-40]. It obviously shows that the $\alpha$-$\kappa$-$\mu$ fading model is more effective and practical than $\alpha$-$\mu$, $\kappa$-$\mu$, Rayleigh, Rice and Nakagamim distributions. However, although a large number of papers have been devoted to study the sensing performance over generalized fading channels, no studies are related to achieve SS over generalized non-linear LOS fading channels $\alpha$-$\kappa$-$\mu$ in the open technical literatures, in addition how to complete and improve the performance evaluations of detection algorithm is a key problem to be solved.

Motivated by above, due to the merits of ED, in this paper we consider ED algorithm over α-κ-µ generalized fading channels to analyze the generic unified detection model. To summarize, the novelty contributions of this paper are summarized as follows:

1) The models of α-κ-µ LOS fading channels have been derived under instantaneous SNR condition to represent the non-linear and small-scale variation of the fading signals, which could be used for investigating the real-time and short-range sensing features in SS under severe fading communication scenarios.

2) The novel and unified exact close-form framework of detection models are derived for different fading scenarios to achieve the SS over α-κ-µ generalized fading channels under instantaneous SNR conditions.

3) To the best of the author`s knowledge, the moment generating function (MGF) method with probability density function-based (PDF-based) approach is firstly utilized to deduce the close-form detective sensing expressions over α-κ-µ fading models.

4) The novel close-form analytical expressions for area under the receiver operating characteristics curve (AUC) and average AUC ($\overline{AUC}$) are derived over α-κ-µ fading channels for revealing and quantifying the relationship between the behavior of ED with the variations of the involved fading parameters algebraically.

5) The offered expressions of AUC and $\overline{AUC}$ from 4) are extended not only limited by integer values, but also for non-integer variables, although the exact computation of AUC and $\overline{AUC}$ are hard to derive.

6) It is shown that if average SNR $\bar{\gamma} \geq 3dB$, the detection performance measurement of ED (with average $\overline{AUC}$) is proportional to the value of nonlinear fading parameter α whenever α is high or low.

7) CSS under diversity reception cases such as MRC, square law combining (SLC) and square Law selection (SLS) have been considered to mitigate the shadowed fading features and improve the corresponding detection probability practically.

8) The tight closed-form upper bound of detection probability with MRC diversity technology is derived to evaluate the detection optimized performance theoretically.

The remaining of this paper is organized as follows. In section II the ED system models are presented. In section III the *p.d.f.* of *α-κ-µ* fading models are derived under instantaneous SNR condition. In Section IV the close-form detection expressions have been derived over generalized fading channels. In Section V the $\overline{AUC}$ performance of ED which is implemented over *α-κ-µ* fading models have been analyzed. The derivation of CSS with diversity reception is given in Section VI. Simulation results are presented in Section VII and conclusions are provided in Section VIII.

## II. SYSTEM MODEL

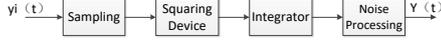

**Fig.1** *ED system model*

The ED model can be assumed to be the binary hypothesis-testing problem in Eq. (1) to determine the absence or presence of unknown wireless signal ($H_0$: signal is absent; $H_1$: signal is present) [12],

$$H_0: y(t) = n(t) \\ H_1: y(t) = h \cdot s(t) + n(t) \quad (1)$$

where $y(t)$ is the received signal, $n(t)$ denotes zero-mean complex additive white Gaussian noise (AWGN), $h$ is the wireless channel gain, $s(t)$ denotes the transmitted primary signal. From Fig.1 the ED model can be obtained,

$$Y = \frac{1}{m}\sum_{i=1}^{m} \frac{y_i^2(t)}{\sigma_i^2} \quad (2)$$

where $m$ is the sampling number of received signal, $\sigma_i^2$ is the AWGN for received signal $y_i(t)$. By defining the time bandwidth product as $u=TW$, $T$ denotes the time interval and $W$ denotes the single-sided signal bandwidth. The probability of detection $P_d$ and the probability of false alarm $P_f$ can be expressed as [14],

$$P_d = P_r(y > \lambda | H_1) = Q_u(\sqrt{2\gamma}, \sqrt{\lambda}) \quad (3)$$

$$P_f = P_r(y > \lambda | H_0) = \frac{\Gamma(u, \frac{\lambda}{2})}{\Gamma(u)} \quad (4)$$

where $Q_u(a,b)$ is the $u$-th order generalized Marcum-$Q$ function, $\Gamma(.)$ is the Gamma function which is defined by the integral $\Gamma(x) = \int_0^\infty t^{x-1} e^{-t} dt$ and $\Gamma(.,.)$ is the incomplete Gamma function which is defined by the integral $\Gamma(a,x) = \int_x^\infty t^{a-1} e^{-t} dt$. $\gamma$ is the instantaneous signal-to-ratio (SNR) and $\lambda$ is the ED threshold [14]. Furthermore, the cumulative density function (*c.d.f.*) of the binary hypothesis-testing in Eq. (1) can be obtained,

$$Y \sim \begin{cases} \chi_{2u}^2; & H_0 \\ \chi_{2u}^2(2\gamma); & H_1 \end{cases} \quad (5)$$

From Eq. (2) the probability density function (*p.d.f.*) of $y(t)$ can be expressed as,

$$f_Y(y) = \begin{cases} \dfrac{1}{2^u \Gamma(\frac{u}{2})} y^{u-1} e^{-\frac{y}{2}}; & H_0 \\ \dfrac{1}{2}(\dfrac{y}{2\gamma})^{\frac{u}{2}-\frac{1}{2}} e^{-\frac{y}{2}-\gamma} I_{u-1}(\sqrt{2y\gamma}); & H_1 \end{cases} \quad (6)$$

where $I_{u-1}(.)$ is the first kind modified Bessel function with the order $u$-1.

## III. THE α-κ-μ FADING MODELS

The α-κ-μ fading distribution is a general fading distribution that can represent the small-scale fading characteristics and describe the received signal which is propagated in the non-linear LOS condition. For the normalized α-κ-μ distribution with envelope $R$, the normalized envelope $r = \sqrt{E(R^2)}$ ($E(.)$ is the expectation and $r$ is the root mean square value) and $y=r^2$, the envelope *p.d.f.* can be shown as [26],

$$f_P^{\alpha-\kappa-\mu}(\rho) = \frac{\alpha\mu\kappa^{\frac{1-\mu}{2}}(1+\kappa)^{\frac{1+\mu}{2}} \rho^{\frac{\alpha(1+\mu)}{2}-1}}{\exp(\kappa\mu + \mu\rho^\alpha + \kappa\mu\rho^\alpha)} \cdot \\ I_{\mu-1}(2\mu\sqrt{\kappa(1+\kappa)}\rho^{\frac{\alpha}{2}}) \quad (7)$$

where parameter α indicates the nonlinear characteristics of the propagation medium, variable κ represents the ratio between the total power of the dominant components and the total power of the scattered waves, parameter μ is related to the number of multipath waves. For the fading signal with the power $w=R^2$ and the normalized power is $w/E(w)$, the power probability density function can be expressed as,

$$f_W^{\alpha-\kappa-\mu}(w) = \frac{\alpha\mu\kappa^{\frac{1-\mu}{2}}(1+\kappa)^{\frac{1+\mu}{2}}}{2\exp(\kappa\mu)} \cdot \frac{w^{\frac{\alpha(1+\mu)}{4}-1}}{\overline{w}^{\frac{\alpha(1+\mu)}{4}}} \cdot$$

$$\exp(-\mu\frac{w^{\frac{\alpha}{2}}}{\overline{w}^{\frac{\alpha}{2}}} - \kappa\mu\frac{w^{\frac{\alpha}{2}}}{\overline{w}^{\frac{\alpha}{2}}}) \cdot I_{\mu-1}(2\mu\sqrt{\kappa(1+\kappa)}\frac{w^{\frac{\alpha}{4}}}{\overline{w}^{\frac{\alpha}{4}}})$$

(8)

Then the *p.d.f.* of the instantaneous SNR $\gamma$ over the $\alpha$-$\kappa$-$\mu$ fading channels can be derived from [Eq. (2), Eq. (6), Eq. (7) and Eq. (8), 27],

$$f_\gamma^{\alpha-\kappa-\mu}(\gamma) = A_1 \cdot \frac{\gamma^{\frac{\alpha(1+\mu)}{4}-1}}{\overline{\gamma}^{\frac{\alpha(1+\mu)}{4}}} \cdot \exp(-\mu\frac{\gamma^{\frac{\alpha}{2}}}{\overline{\gamma}^{\frac{\alpha}{2}}} - \kappa\mu\frac{\gamma^{\frac{\alpha}{2}}}{\overline{\gamma}^{\frac{\alpha}{2}}}) \cdot$$

$$I_{\mu-1}(2\mu\sqrt{\kappa(1+\kappa)}\frac{\gamma^{\frac{\alpha}{4}}}{\overline{\gamma}^{\frac{\alpha}{4}}})$$

(9)

where

$$A_1 = \frac{\alpha\mu\kappa^{\frac{1-\mu}{2}}(1+\kappa)^{\frac{1+\mu}{2}}}{2\exp(\kappa\mu)}$$ (10)

## IV. SPECTRUM SENSING OVER FADING CHANNELS

Evaluating the detection probability $P_d$ of ED algorithm with Eq. (3) and Marcum-$Q$ function from Ref. [15] and Ref. [16] for single sensing node,

$$Q_u^{ED}(\sqrt{2\gamma},\sqrt{\lambda}) = \sum_{l=0}^{\infty}\frac{\exp(-\gamma)\gamma^l \Gamma(l+u,\frac{\lambda}{2})}{\Gamma(l+1)\Gamma(l+u)}$$ (11)

It is recalled that the average probability of detection over $\alpha$-$\kappa$-$\mu$ generalized fading channels can be obtained as,

$$\overline{P_d^{\alpha-\kappa-\mu}} = \int_0^\infty P_d^{ED} \cdot f_\gamma^{\alpha-\kappa-\mu}(\gamma)d\gamma$$ (12)

where $P_d^{ED}$ denotes the probability of detection, $f_\gamma^{\alpha-\kappa-\mu}(\gamma)$ is the *p.d.f.* of the $\alpha$-$\kappa$-$\mu$ distribution under instantaneous SNR condition. Then by substituting Eq. (9) and Eq. (11), the average probability of detection $\overline{P_d^{\alpha-\kappa-\mu}}$ can be expressed as,

$$\overline{P_d^{\alpha-\kappa-\mu}} = \int_0^\infty \sum_{l=0}^{\infty}\frac{\exp(-\gamma)\gamma^l \Gamma(l+u,\frac{\lambda}{2})}{\Gamma(l+1)\Gamma(l+u)} \cdot$$

$$f_\gamma^{\alpha-\kappa-\mu}(\gamma)d\gamma$$
(13)

Adopting the moment generating function (*MGF*) algorithm, the *p.d.f.* of *MGF* under the instantaneous SNR over $\alpha$-$\kappa$-$\mu$ fading channels is shown as [41],

$$\phi_\gamma^{\alpha-\kappa-\mu}(s) = E^{\alpha-\kappa-\mu}(e^{-s\gamma})$$ (14)
$$= \int_0^\infty e^{-s\gamma} \cdot f_\gamma^{\alpha-\kappa-\mu}(\gamma)d\gamma$$

where $E(.)$ denotes the expectation. Deducing the *n*-th derivative of the *MGF* as,

$$\frac{\delta^{(n)}\phi_\gamma^{\alpha-\kappa-\mu}(s)}{\delta s^n} = \frac{\delta^{(n)}E^{\alpha-\kappa-\mu}(e^{-s\gamma})}{\delta s^n}$$

$$= \int_0^\infty (-\gamma)^n e^{-s\gamma} \cdot f_\gamma^{\alpha-\kappa-\mu}(\gamma)d\gamma$$ (15)

Likewise, with Eq. (15) the generic expression $\overline{P_d^{\alpha-\kappa-\mu}}$ can be obtained in another way,

$$\overline{P_d^{\alpha-\kappa-\mu}} = \sum_{l=0}^{\infty}\frac{(-1)^n \Gamma(l+u,\frac{\lambda}{2})}{\Gamma(l+1)\Gamma(l+u)} \cdot \frac{\delta^{(l)}\phi_\gamma^{\alpha-\kappa-\mu}(s)}{\delta s^l}\bigg|_{s=1}$$

(16)

Moreover, the closed-form expression over $\alpha$-$\kappa$-$\mu$ fading channels can be evaluated as,

$$\phi_\gamma^{(n),\alpha-\kappa-\mu}(s) = \int_0^\infty (-\gamma)^n e^{-s\gamma} \cdot A_1 \cdot \frac{\gamma^{\frac{\alpha(1+\mu)}{4}-1}}{\overline{\gamma}^{\frac{\alpha(1+\mu)}{4}}} \cdot$$

$$\exp(-\mu\frac{\gamma^{\frac{\alpha}{2}}}{\overline{\gamma}^{\frac{\alpha}{2}}} - \kappa\mu\frac{\gamma^{\frac{\alpha}{2}}}{\overline{\gamma}^{\frac{\alpha}{2}}}) \cdot I_{\mu-1}(2\mu\sqrt{\kappa(1+\kappa)}\frac{\gamma^{\frac{\alpha}{4}}}{\overline{\gamma}^{\frac{\alpha}{4}}})d\gamma$$

(17)

The modified Bessel function of the first kind with the order $v$ in Eq. (17) can be simplified with [Eq. (8.445), 43]

$$I_v(z) = \sum_{k=0}^{\infty}\frac{1}{\Gamma(k+1)\Gamma(v+k+1)} \cdot (\frac{z}{2})^{2k+v}$$ (18)

Then Eq. (17) can be derived by using infinite series representation,

$$\phi_{\gamma}^{(n),\alpha-\kappa-\mu}(s) = (-1)^n A_1 \cdot$$

$$\sum_{k=0}^{\infty} \frac{\kappa^{k+\frac{\mu-1}{2}}(\kappa+1)^{k+\frac{\mu-1}{2}}\mu^{2k+\mu-1}}{\Gamma(k+1)\Gamma(k+\mu)\gamma^{-\frac{\alpha}{4}(2k+2\mu)}} \cdot \int_0^{\infty} \frac{(s\gamma)^{n+\frac{\alpha}{4}(2k+2\mu)-1}}{s^{n+\frac{\alpha}{4}(2k+2\mu)}} \cdot$$

$$\exp(-s\gamma - \frac{\mu}{\gamma^{-\frac{\alpha}{2}}s^{\frac{\alpha}{2}}}(s\gamma)^{\frac{\alpha}{2}} - \frac{\kappa\mu}{\gamma^{-\frac{\alpha}{2}}s^{\frac{\alpha}{2}}}(s\gamma)^{\frac{\alpha}{2}})d(s\gamma)$$

(19)

Evaluating the integral in Eq. (19) with extended incomplete Gamma function in Ref. [44] and Ref. [45],

$$\Gamma(\alpha,x,b,\beta) = \int_x^{\infty} t^{\alpha-1} e^{-t-bt^{-\beta}} dt \quad (20)$$

Next, the closed-form expression of Eq. (19) can be simplified as,

$$\phi_{\gamma}^{(n),\alpha-\kappa-\mu}(s) = (-1)^n A_1 \cdot$$

$$\sum_{k=0}^{\infty} \frac{\kappa^{k+\frac{\mu-1}{2}}(\kappa+1)^{k+\frac{\mu-1}{2}}\mu^{2k+\mu-1}}{\Gamma(k+1)\Gamma(k+\mu)\gamma^{-\frac{\alpha}{4}(2k+2\mu)}s^{n+\frac{\alpha}{4}(2k+2\mu)}} \cdot \quad (21)$$

$$\Gamma(n+\frac{\alpha}{4}(2k+2\mu),0,\frac{\mu}{\gamma^{-\frac{\alpha}{2}}s^{\frac{\alpha}{2}}}+\frac{\kappa\mu}{\gamma^{-\frac{\alpha}{2}}s^{\frac{\alpha}{2}}},-\frac{\alpha}{2})$$

Furthermore, the Eq. (21) can be evaluated with [Theorem 3.1, 45] as,

$$\phi_{\gamma}^{(n),\alpha-\kappa-\mu}(s) = (-1)^n A_1 \cdot A_2$$

$$\sum_{k=0}^{\infty} \frac{\kappa^k (\kappa+1)^k \mu^{2k}}{\Gamma(k+1)\Gamma(k+\mu)\gamma^{-\alpha k} s^{\alpha k}} \cdot$$

$$H_{0,2}^{2,0}[\frac{\mu}{\gamma^{-\frac{\alpha}{2}}s^{\frac{\alpha}{2}}}+\frac{\kappa\mu}{\gamma^{-\frac{\alpha}{2}}s^{\frac{\alpha}{2}}} \mid \begin{array}{c} -, - \\ (0,1), (n+\frac{\alpha}{4}(2k+2\mu), -\frac{\alpha}{2}) \end{array}]$$

(22)

where

$$A_2 = \frac{\kappa^{\frac{\mu-1}{2}}(\kappa+1)^{\frac{\mu-1}{2}}\mu^{\mu-1}}{\gamma^{-\alpha\mu}s^{n+\alpha\mu}} \quad (23)$$

The close-form expression of the FOX-H function can be expressed as,

$$H_{p,q}^{m,n}[z \mid \begin{array}{c} (a_1,A_1),(a_2,A_2),...,(a_p,A_p) \\ (b_1,B_1),(b_2,B_2),...,(b_q,B_q) \end{array}] =$$

$$\frac{1}{2\pi i}\int_L \frac{(\prod_{j=1}^m \Gamma(b_j+B_js))(\prod_{j=1}^n \Gamma(1-a_j-A_js))}{(\prod_{j=m+1}^q \Gamma(1-b_j-B_js))(\prod_{j=n+1}^p \Gamma(a_j+A_js))} z^{-s} ds$$

(24)

## V. AVERAGE AREA UNDER THE ROC CURVE (AUC) OVER α-κ-μ FADING CHANNELS

### 5.1 AUC under instantaneous SNR condition

The AUC measurement is usually used for characterizing the performance of ED [46], health care field tests [47] and machine learning algorithms [48] such as plotting $P_d$ versus $P_f$ (ROC) or missed detection probability $P_m$ ($P_m=1-P_d$) versus $P_f$ (complementary ROC). It varies between 0.5 and 1, and it can be considered comprehensively by analyzing $P_d$ and $P_f$ which represents the probability of more correct choosing, so it makes sense that deriving the close-form expressions for the AUC of ED under appropriate conditions to quantify the detection performance like Ref. [49-52]. Here we introduce the ED threshold $\lambda$ varies from 0 to ∞ to analyze the capability of energy detector. When the instantaneous SNR value denotes $\gamma$, the AUC can be shown as [52],

$$AUC(\gamma) = \int_0^{\infty} P_d(\gamma,\lambda) dP_f(u,\lambda) \quad (25)$$

Taking the derivative with $P_f(u,\lambda)$, Eq. (25) can be written as,

$$AUC(\gamma) = -\int_0^{\infty} P_d(\gamma,\lambda) \frac{\partial P_f(u,\lambda)}{\partial \lambda} d\lambda \quad (26)$$

With Eq. (3), Eq. (4) and [Eq. (12), 52], Eq. (26) can be derived as,

$$AUC(\gamma) = 1 - \sum_{l=0}^{u-1}\sum_{i=0}^{l}\binom{l+u-1}{l-i}\int_0^{\infty}\frac{\gamma^i \exp(-\frac{\gamma}{2})}{2^{l+i+u}\cdot i!}d\gamma$$

(27)

### 5.2 $\overline{AUC}$ over α-κ-μ fading channels

The average AUC ($\overline{AUC}$) can be investigated

with the *p.d.f.* of generalized fading models for the instantaneous SNR distribution to indicate the properties of the fading channels, therefore, $\overline{AUC}$ can be shown as,

$$\overline{AUC} = \int_0^\infty AUC(\gamma) \cdot f_\gamma^{\alpha-\kappa-\mu}(\gamma) d\gamma \qquad (28)$$

where

$$\binom{a}{b} = \frac{a!}{b!(a-b)!} \qquad (29)$$

From Ref. [21], Eq. (28) can be derived with the help of Eq. (25)-Eq. (27),

$$\overline{AUC} = 1 - \sum_{l=0}^{u-1}\sum_{i=0}^{l}\binom{l+u-1}{l-i}\frac{1}{2^{l+i+u}\cdot i!}\int_0^\infty \gamma^i \cdot \exp(-\frac{\gamma}{2}) f_\gamma^{\alpha-\kappa-\mu}(\gamma) d\gamma \qquad (30)$$

Depending on Eq. (9) and Eq. (30), Eq. (30) can be derived as,

$$\overline{AUC} = 1 - A_1 \cdot \sum_{l=0}^{u-1}\sum_{i=0}^{l}\binom{l+u-1}{l-i}\frac{1}{2^{l+i+u}\cdot i!} \cdot$$
$$\int_0^\infty \frac{\gamma^{i+\frac{\alpha(1+\mu)}{4}-1}}{\overline{\gamma}^{\frac{\alpha(1+\mu)}{4}}} \cdot \exp(-\frac{\gamma}{2} - \mu\frac{\gamma^{\frac{\alpha}{2}}}{\overline{\gamma}^{\frac{\alpha}{2}}} - \kappa\mu\frac{\gamma^{\frac{\alpha}{2}}}{\overline{\gamma}^{\frac{\alpha}{2}}}) \cdot \qquad (31)$$
$$I_{\mu-1}(2\mu\sqrt{\kappa(1+\kappa)}\frac{\gamma^{\frac{\alpha}{4}}}{\overline{\gamma}^{\frac{\alpha}{4}}}) d\gamma$$

Like Eq. (17)-Eq. (19), by means of [Eq. (8.445), 43], then Eq. (31) can be derived as,

$$\overline{AUC} = 1 - A_1 \cdot \sum_{l=0}^{u-1}\sum_{i=0}^{l}\binom{l+u-1}{l-i}\frac{1}{2^{l+i+u}\cdot i!} \cdot$$
$$\sum_{k=0}^{\infty}\frac{\mu^{2k+\mu-1}\kappa^{k+\frac{\mu-1}{2}}(1+\kappa)^{k+\frac{\mu-1}{2}}}{k!\Gamma(k+\mu)}\int_0^\infty \frac{\gamma^{i+\frac{\alpha\mu}{2}+\frac{\alpha k}{2}-1}}{\overline{\gamma}^{\frac{\alpha\mu}{2}+\frac{\alpha k}{2}}} \cdot \qquad (32)$$
$$\exp(-\frac{\gamma}{2} - \mu\frac{\gamma^{\frac{\alpha}{2}}}{\overline{\gamma}^{\frac{\alpha}{2}}} - \kappa\mu\frac{\gamma^{\frac{\alpha}{2}}}{\overline{\gamma}^{\frac{\alpha}{2}}}) d\gamma$$

Next, with the aid of Eq. (10), Eq. (32) can be simplified as,

$$\overline{AUC} = 1 - \sum_{k=0}^{\infty}\sum_{l=0}^{u-1}\sum_{i=0}^{l}\binom{l+u-1}{l-i}\frac{1}{2^{l+i+u}\cdot i!} \cdot$$
$$\frac{\alpha\mu^{2k+\mu}\kappa^k(1+\kappa)^{k+\mu}}{\exp(\kappa\mu)k!\Gamma(k+\mu)} \cdot \qquad (33)$$
$$\underbrace{\int_0^\infty \frac{\gamma^{i+\frac{\alpha\mu}{2}+\frac{\alpha k}{2}-1}}{\overline{\gamma}^{\frac{\alpha\mu}{2}+\frac{\alpha k}{2}}} \cdot \exp(-\frac{\gamma}{2} - \mu\frac{\gamma^{\frac{\alpha}{2}}}{\overline{\gamma}^{\frac{\alpha}{2}}} - \kappa\mu\frac{\gamma^{\frac{\alpha}{2}}}{\overline{\gamma}^{\frac{\alpha}{2}}}) d\gamma}_{B_1}$$

Simplifying the integral $B_1$ in Eq. (33) based on Taylor series [43],

$$B_1 = \int_0^\infty \frac{\gamma^{i+\frac{\alpha\mu}{2}+\frac{\alpha k}{2}-1}}{\overline{\gamma}^{\frac{\alpha\mu}{2}+\frac{\alpha k}{2}}} \cdot \sum_{n=0}^{\infty}\frac{(-\frac{\gamma}{2}-\mu\frac{\gamma^{\frac{\alpha}{2}}}{\overline{\gamma}^{\frac{\alpha}{2}}} - \kappa\mu\frac{\gamma^{\frac{\alpha}{2}}}{\overline{\gamma}^{\frac{\alpha}{2}}})^n}{n!} d\gamma \qquad (34)$$

Expanding Eq. (34) by the Binomial theorem [43],

$$B_1 = \int_0^\infty \frac{\gamma^{i+\frac{\alpha\mu}{2}+\frac{\alpha k}{2}-1}}{\overline{\gamma}^{\frac{\alpha\mu}{2}+\frac{\alpha k}{2}}} \cdot \sum_{n=0}^{\infty}\sum_{w=0}^{\infty}\frac{(-1)^n}{n!}\binom{n}{w}\cdot(\frac{\mu+\kappa\mu}{\overline{\gamma}^{\frac{\alpha}{2}}})^w \cdot$$
$$\gamma^{n+w(\frac{\alpha}{2}-1)} d\gamma \qquad (35)$$

Next, the integral in Eq. (35) can be simplified as,

$$B_1 = \sum_{n=0}^{\infty}\sum_{w=0}^{n}\frac{(-1)^n}{n!}\binom{n}{w}\cdot(\frac{\mu+\kappa\mu}{\overline{\gamma}^{\frac{\alpha}{2}}})^w \cdot$$
$$\frac{\gamma^{n+i+\frac{\alpha\mu}{2}+\frac{\alpha k}{2}+w(\frac{\alpha}{2}-1)}\Big|_{0^+}^{\infty}}{\overline{\gamma}^{\frac{\alpha\mu}{2}+\frac{\alpha k}{2}} \cdot (n+i+\frac{\alpha\mu}{2}+\frac{\alpha k}{2}+w(\frac{\alpha}{2}-1))} \qquad (36)$$

Lastly, the exact close-form expression can be obtained by substituting Eq. (36) into Eq. (33), Eq. (33) can be simplified as,

$$\overline{AUC} = 1 - A_3 \sum_{k=0}^{\infty}\sum_{l=0}^{u-1}\sum_{i=0}^{l}\sum_{n=0}^{\infty}\sum_{w=0}^{n}\binom{l+u-1}{l-i}\cdot\frac{1}{2^{l+i+u}\cdot i!} \cdot$$
$$\frac{\mu^{2k}\kappa^k(1+\kappa)^k}{\Gamma(k+1)\Gamma(k+\mu)} \cdot \frac{(-1)^n(\mu+\kappa\mu)^w}{\Gamma(w+1)\Gamma(n-w+1)} \cdot$$
$$\frac{\gamma^{n+i+\frac{\alpha\mu}{2}+\frac{\alpha k}{2}+w(\frac{\alpha}{2}-1)}\Big|_{0^+}^{\infty}}{\overline{\gamma}^{\frac{\alpha k}{2}+\frac{\alpha w}{2}} \cdot (n+i+\frac{\alpha\mu}{2}+\frac{\alpha k}{2}+w(\frac{\alpha}{2}-1))}$$

(37)

where

$$A_3 = \frac{\alpha\mu^\mu(1+\kappa)^\mu}{\exp(\kappa\mu)\bar{\gamma}^{-\frac{\alpha\mu}{2}}} \quad (38)$$

## VI. ED WITH RECEIVER DIVERSITY OVER α-κ-μ FADING CHANNELS

### 6.1 Upper bound of detection performance with MRC

It makes sense that implementing ED with diversity over fading channels at the receiver can increase the receiver SNR for the $L$ diversity paths which are normally supposed to independent and identically distribution (*i.i.d.*) [41,42].

For MRC, at the receiving end the total output SNR will be combined by all diversity branches and the collected data is added up to the reserved energy detector. Although MRC requires prior channel knowledge of the signal, in practice it estimates the upper bound on the performance of ED [11]. The total instantaneous SNR of MRC is given by,

$$\gamma_{MRC} = \sum_{i=1}^{L}\gamma_i \quad (39)$$

where $L$ denotes the number of antennas for each SU, the SNR of *i*-th receiver branch is defined as $\gamma_i$.

When the number of diversity branches at the receiver is $L$, the average detection probability with MRC can be computed by substituting Eq. (39) and Eq. (22) in Eq. (13),

$$\overline{P_{d-MRC}^{\alpha-\kappa-\mu}} = \sum_{l=0}^{\infty}\frac{(-1)^l\Gamma(l+u,\frac{\lambda}{2})}{\Gamma(l+1)\Gamma(l+u)} \cdot \frac{\delta^{(l)}\phi_{\gamma_{MRC}=\sum_{i=1}^{L}\gamma_i}^{\alpha-\kappa-\mu}(s)}{\delta s^l}\bigg|_{s=1} \quad (40)$$

where the $\phi_{\gamma_{MRC}=\sum_{i=1}^{L}\gamma_i}^{\alpha-\kappa-\mu}(s)$ denotes the *MGF* with the total SNR of diversity branches. Using [Eq. (24), 21], the *MGF* based on MRC scheme can be evaluated as,

$$\phi_{\gamma_{MRC}=\sum_{i=1}^{L}\gamma_i}^{\alpha-\kappa-\mu}(s) = \prod_{i=1}^{L}\phi_{\gamma_i}^{\alpha-\kappa-\mu}(s) \quad (41)$$

On the other hand, it is worth noting that the Leibniz's rule can be obtained when the number of product terms of functions is two [Eq. (0.42), 43],

$$\frac{d^{(n)}(u(x)\cdot v(x))}{dx^n} = \sum_{i=0}^{n}\binom{n}{i}\frac{d^{(i)}u(x)}{dx^i}\frac{d^{(n-i)}v(x)}{dx^{(n-i)}} \quad (42)$$

Deriving the *n*-order Leibniz's rule with the aid of [Eq. (26), 21], the *n*-th derivative of (41) can be deduced by means of Eq. (14), Eq. (15) and Eq. (42),

$$\phi_{\gamma_{MRC}=\sum_{i=1}^{L}\gamma_i}^{(n),\alpha-\kappa-\mu}(s) = \sum_{n_1=0}^{n}\sum_{n_2=0}^{n_1}\cdots\sum_{n_{L-1}=0}^{n_{L-2}}\binom{n}{n_1}\binom{n_1}{n_2}\cdots\binom{n_{L-2}}{n_{L-1}}\cdot$$
$$\phi_{\gamma_1}^{(n-n_1)}(s)\phi_{\gamma_2}^{(n_1-n_2)}(s)\cdots\phi_{\gamma_{L-1}}^{(n_{L-2}-n_{L-1})}(s)\phi_{\gamma_L}^{(n_{L-1})}(s) \quad (43)$$

Furthermore, from Eq. (22), Eq. (24), Eq. (40) and Eq. (43), the close-form detective expressions of ED over α-κ-μ fading channels are given by,

$$\overline{P_{d-MRC}^{\alpha-\kappa-\mu}} = \sum_{l=0}^{\infty}\frac{(-1)^l\Gamma(l+u,\frac{\lambda}{2})}{\Gamma(l+1)\Gamma(l+u)}\cdot$$
$$\sum_{n_1=0}^{l}\sum_{n_2=0}^{n_1}\cdots\sum_{n_{L-1}=0}^{n_{L-2}}\binom{l}{n_1}\binom{n_1}{n_2}\cdots\binom{n_{L-2}}{n_{L-1}}\cdot$$
$$\phi_{\gamma_1}^{(l-n_1)}(s)\phi_{\gamma_2}^{(n_1-n_2)}(s)\cdots\phi_{\gamma_{L-1}}^{(n_{L-2}-n_{L-1})}(s)\phi_{\gamma_L}^{(n_{L-1})}(s)\bigg|_{s=1} \quad (44)$$

where

$$\phi_\gamma^{(n),\alpha-\kappa-\mu}(s) = (-1)^n A_1 \cdot A_2$$
$$\sum_{k=0}^{\infty}\frac{\kappa^k(\kappa+1)^k\mu^{2k}}{\Gamma(k+1)\Gamma(k+\mu)\bar{\gamma}^{-\alpha k}s^{\alpha k}}\cdot$$
$$H_{0,2}^{2,0}\left[\frac{\mu}{\bar{\gamma}^{-\frac{\alpha}{2}}s^{\frac{\alpha}{2}}}+\frac{\kappa\mu}{\bar{\gamma}^{-\frac{\alpha}{2}}s^{\frac{\alpha}{2}}}\bigg|\begin{array}{c}-,-\\(0,1),(n+\frac{\alpha}{4}(2k+2\mu),-\frac{\alpha}{2})\end{array}\right] \quad (45)$$

### 6.2 Cooperative spectrum sensing with diversity reception SLC and SLS

Square Law Combining (SLC): Similar to MRC, the SLC scheme requires the received signals are integrated and squared, further the output signals are summed together [51]. The decision model follows a central chi-square distribution with $2Lu$ degrees of freedom if the binary hypothesis testing is $H_0$, or a non-central chi-square distribution with $2Lu$ degrees of freedom under binary hypothesis testing $H_1$. The total received SNR $\gamma_{SLC}$ is equal to combined instantaneous SNR $\gamma_{MRC}$ under MRC, besides the time bandwidth product $u$ is replaced by $Lu$ for Eq. (44) to represent the detection capacity over $\alpha$-$\kappa$-$\mu$ fading channels. When the number of diversity branches is $L$, the detection probability can be shown as,

$$\overline{P_{d-SLC}^{\alpha-\kappa-\mu}} = \sum_{l=0}^{\infty} \frac{(-1)^l \Gamma(l+Lu, \frac{\lambda}{2})}{\Gamma(l+1)\Gamma(l+Lu)} \cdot$$
$$\sum_{n_1=0}^{l} \sum_{n_2=0}^{n_1} \cdots \sum_{n_{L-1}=0}^{n_{L-2}} \binom{l}{n_1}\binom{n_1}{n_2}\cdots\binom{n_{L-2}}{n_{L-1}} \cdot$$
$$\phi_{\gamma_1}^{(l-n_1)}(s) \phi_{\gamma_2}^{(n_1-n_2)}(s) \cdots \phi_{\gamma_{L-1}}^{(n_{L-2}-n_{L-1})}(s) \phi_{\gamma_L}^{(n_{L-1})}(s)\big|_{s=1}$$

(46)

where

$$\gamma_{SLC} = \gamma_{MRC} \qquad (47)$$

Square Law selection (SLS): In SLS scheme the maximum decision statistics $y_{SLS}$ ($y_{SLS} = \max\{y_1, y_2, ..., y_L\}$) is selected to calculate the average probability of detection of ED over fading channels [53]. The detection probability can be expressed as,

$$\overline{P_{d-SLS}^{\alpha-\kappa-\mu}} = 1 - \prod_{i=1}^{L}(1-P_{d-SLS}) \qquad (48)$$

where

$$P_{d-SLS} = (-1)^n \frac{\alpha\mu\kappa^{\frac{1-\mu}{2}}(1+\kappa)^{\frac{1+\mu}{2}}}{2\exp(\kappa\mu)} \cdot$$
$$\frac{\kappa^{\frac{\mu-1}{2}}(\kappa+1)^{\frac{\mu-1}{2}}\mu^{\mu-1}}{\overline{\gamma}^{-\alpha\mu} s^{n+\alpha\mu}} \cdot \sum_{k=0}^{\infty} \frac{\kappa^k(\kappa+1)^k \mu^{2k}}{\Gamma(k+1)\Gamma(k+\mu)\overline{\gamma}^{-\alpha k} s^{\alpha k}} \cdot$$
$$H_{0,2}^{2,0}\left[\frac{\mu}{\overline{\gamma}^{-\frac{\alpha}{2}} s^{\frac{\alpha}{2}}} + \frac{\kappa\mu}{\overline{\gamma}^{-\frac{\alpha}{2}} s^{\frac{\alpha}{2}}} \bigg| \begin{matrix}-,-\\(0,1),(n+\frac{\alpha}{4}(2k+2\mu),-\frac{\alpha}{2})\end{matrix}\right]$$

(49)

Cooperative spectrum sensing (CSS): CSS have been proposed to help the shadowed SUs detect the occupied wideband of PUs and improve the sensing capability for severe multipath fading, in which firstly SUs send their own decisions to the fusion center (FC) respectively, then FC makes the global decision by combining the received information to determine the absence or presence of PU [24]. In view of the above, the CSS decision rule follows,

$$D = \sum_{i=1}^{N} D_i = \begin{cases} <n, & H_0 \\ \geq n, & H_1 \end{cases} \qquad (50)$$

where $D$ is the sum of sensed decisions with hard decision for sending "1-bit" under $H_1$ or "0-bit" under $H_0$, $N$ is the number of collaborative users. From Eq. (50) the AND-rule is corresponding to the case of $n=N$ and OR-rule is corresponding to the case of $n=1$, otherwise it represents the "$n$-out-of-$N$" collaborative voting rule for $n\neq 1$ and $n\neq N$. In order to evaluate the detection probability $P_{d\text{-}CSS}$ and the false-alarm probability $P_{f\text{-}CSS}$ of CSS with SLC and SLS for the OR-rule, it can be respectively determined by,

$$P_{d-CSS}^{SLC,\alpha-\kappa-\mu} = 1 - \prod_{i=1}^{N}(1-\overline{P_{d-SLC}^{\alpha-\kappa-\mu}}) \qquad (51)$$

$$P_{d-CSS}^{SLS,\alpha-\kappa-\mu} = 1 - \prod_{i=1}^{N}(1-\overline{P_{d-SLS}^{\alpha-\kappa-\mu}}) \qquad (52)$$

## VII. NUMERICAL SIMULATION AND ANALYSIS

In this section, numerical simulation and analysis for the behavior of ED have been provided to reveal the crucial impact on the performance of SS over $\alpha$-$\kappa$-$\mu$ fading channels with software packages as MATLAB and MATHEMATICA [54]. The corresponding performances of Section IV - Section VI are quantified in the following figures under different fading environments to show the various numerical features of severe shadowing conditions.

Fig.2 illustrates the average detection probability $\overline{P_d^{\alpha-\kappa-\mu}}$ versus average SNR $\bar{\gamma}$ over $\alpha$-$\kappa$-$\mu$ fading channels for different channel parameters $\kappa$ and $\mu$ under constant nonlinear coefficient $\alpha$, when $u=2$ and $P_f=0.01$. It can be observed that although nonlinearity coefficient of channel condition is low, the raise of $\kappa$ and $\mu$ can improve the detection capacity of ED because the higher ratio between the total power of the dominant components and the total power of the scattered waves and more related variable of multipath clusters will lead to more received power of dominant components, when $\alpha=1.35$, $\kappa=0.7$-$1.0$ and $\mu=0.7$-$1.0$. Besides the $\overline{P_d^{\alpha-\kappa-\mu}}$ also improves substantially as $\bar{\gamma}$ raises for low value of constant $\alpha$, and if $\kappa$ and $\mu$ are constant, higher $\alpha$ will lead to higher average detection probability. Furthermore, if $\kappa$ and $\mu$ are invariable, the detection probability improves more quickly for higher $\alpha$ as $\bar{\gamma}$ increases, for example, when $\alpha=1.35$-$1.75$, $\kappa=1.0$ and $\mu=1.0$.

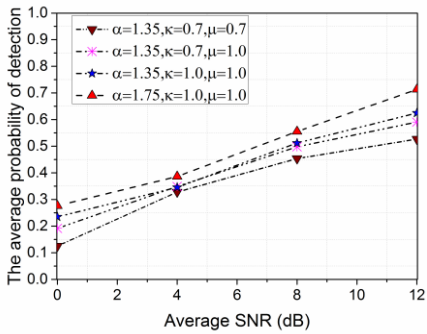

**Fig.2** *Simulation for average probability of detection versus average SNR (dB) with $u=2$ and $P_f=0.01$*

On the other hand, Fig.3 and Fig.4 collectively show the $\overline{AUC}$ value against $\bar{\gamma}$ for different variables of $\alpha$, $\kappa$ and $\mu$ to present corresponding performance characteristics of ED like Ref. [26, 37-40].

Fig.3 depicts $\overline{AUC}$ versus $\bar{\gamma}$ with $u=2$ for low linearity coefficient. If $\kappa=0.7$, $\mu=0.7$ and when $\alpha=1.5$ is twice more than $\alpha=0.7$, the case for $\alpha=1.5$ exceeds the case for $\alpha=0.7$ obviously as $\bar{\gamma}$ increases when $\bar{\gamma}\geq 2dB$. Furthermore, it can be seen that under low values of $\alpha$ small variations of $\mu$ will evidently improve the sensing performance of ED, comparing with the values change for $\kappa$. It has been demonstrated that the effects of related variable of multipath clusters are more critical than the ratio of the total dominant components power to the total scattered waves power if the value of $\alpha$ is low.

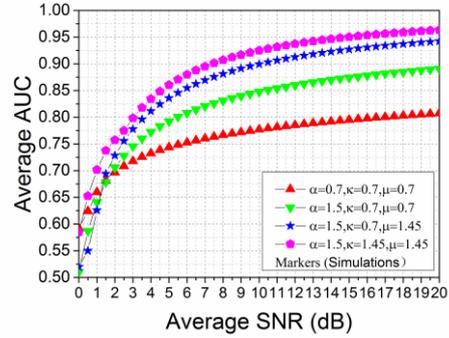

**Fig.3** *Average AUC versus average SNR (dB) over $\alpha$-$\kappa$-$\mu$ fading channels with $u=2$ for low values of $\alpha$*

Fig.4 illustrates $\overline{AUC}$ against average SNR with time bandwidth product $u=2$ for higher nonlinear coefficient compared to $\alpha$ in Fig.3. Similarly higher average SNR is corresponding to better ROC curve performance for ED-based SS scheme. However, although under high values of $\alpha$, it can be seen that higher $\alpha$ will lead to better detection capability, the variations of other channel parameters $\kappa$ and $\mu$ could not significantly alter the $\overline{AUC}$ as parameter $\alpha$ increases. Besides, it can be obtained from Fig.3 and Fig.4 at the same time that higher $\alpha$ corresponds to better simulation results when average SNR is greater than 3 dB, and better $\overline{AUC}$ for higher $\kappa$ and $\mu$ with the same value of $\alpha$ when average SNR is less than 3dB.

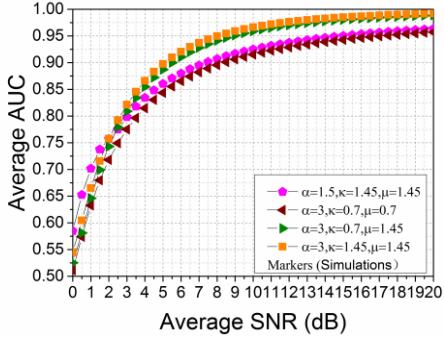

**Fig.4** *Average AUC versus average SNR (dB) over $\alpha$-$\kappa$-$\mu$ fading channels with u=2 for high values of $\alpha$*

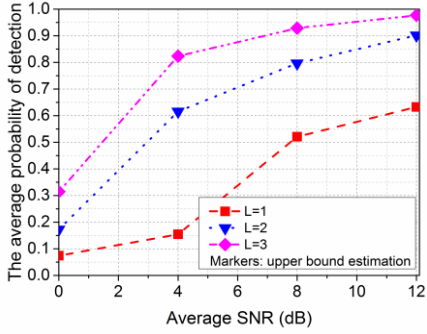

**Fig.5** *The average probability of detection versus average SNR (dB) for upper bound of detection probability with MRC when u=2, $\alpha$=1.35, $\kappa$=1.0, $\mu$=1.0 and $P_f$=0.01*

In Fig.5 the upper bound of detection probability for MRC have been analyzed theoretically with $u=2$ and $P_f$=0.01 when channel parameters are assumed as, $\alpha$=1.35, $\kappa$=1.0 and $\mu$=1.0. It is shown that the performance of detection is proportional to the average SNR and the number of diversity branch. Furthermore, as the diversity branch increases, the average probability of detection can be obviously raised. For example, despite of low $P_f$, if $L$=2, $P_d$≈0.767 which is compared to $P_d$≈0.519 for $L$=1 when average SNR is 8 dB. Likewise, $P_d$≈0.835 for $L$=3 and it far exceeds the detection probability ($P_d$≈0.617) for $L$=2 when average SNR is 4 dB.

On the other side, Fig. 6 and Fig.7 show that CSS with diversity combining can availably improve the sensing performance of ED over $\alpha$-$\kappa$-$\mu$ fading channels when $u$=2, $\alpha$=1.35, $\kappa$=1.0 and $\mu$=1.0. The average probability of detection versus average SNR with the number of collaborative users $N$=1,2 and $N$=2,3 are analyzed respectively to present the relationship between the detection capability and average instantaneous SNR for various cooperative users. It can be indicated that under low fixed values of nonlinear parameter, in practical sensing course the diversity technique and CSS both improve the detection performance, although the $P_f$ for SLC can be expressed as $\Gamma(Lu, \lambda/2)/\Gamma(Lu)$ and the $P_f$ for SLS can be represented as $1-(1-\Gamma(u, \lambda/2)/\Gamma(u))^N$. Moreover, it can be inferred that SLC has better effects for detection probability than SLS under the same channel condition, and the improvement capability for diversity combining is almost similar to CSS.

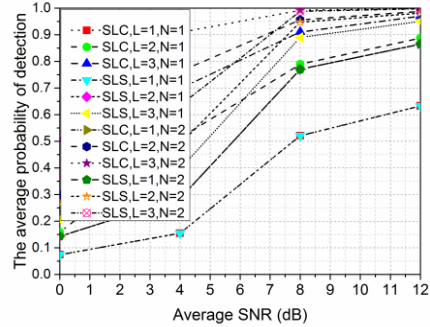

**Fig.6** *The average probability of detection versus average SNR (dB) for CSS with SLC and SLS when u=2, N=1, 2, $\alpha$=1.35, $\kappa$=1.0, $\mu$=1.0 and $P_f$=0.01*

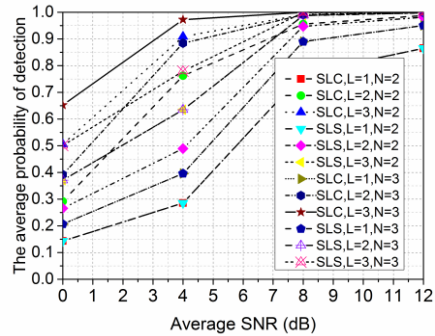

**Fig.7** *The average probability of detection versus average SNR (dB) for CSS with SLC and SLS when u=2, N=2, 3, α=1.35, κ=1.0, μ=1.0 and $P_f$ =0.01*

## VIII. CONCLUSIONS

This paper investigates the SS based on ED that is implemented over *α-κ-μ* fading channels to reveal the relationship between the performance of ED and nonlinear LOS fading channels. The *α-κ-μ* fading models under instantaneous SNR condition have been derived to achieve SS and the novel unified close-form expressions of ED over fading channels have been deduced to show essential sensing probability with *PDF*-based and *MGF* algorithm. In addition, novel exact close-form expressions of average *AUC* have been derived to quantify the behavior of ED under different values of nonlinear coefficient and other crucial fading parameters. Besides, it is demonstrated that diversity technique and CSS can jointly improve the detection performance under severe fading conditions and upper bound with MRC have been inferred to evaluate the sensing performance theoretically. Generally speaking, the sufficient results that are derived above can be completely used to quantify the performance of SS over *α-κ-μ* nonlinear LOS fading scenarios, and it can radically improve the energy efficiency for CR systems in wireless communications.